# Broad Distribution of Stick-slip Events in Slowly Sheared Granular Media: Table-top Production of a Gutenberg-Richter-like Distribution

Michael Bretz[1*], Russell Zaretzki[1,2], Stuart Field[1,3], Namiko Mitarai[4**], and Franco Nori[1,4]

[1] *Department of Physics, University of Michigan, Ann Arbor, Michigan 48109, USA*
[2] *Department of Statistics, University of Tennessee, Knoxville, TN 37996, USA*
[3] *Department of Physics, Colorado State University, Fort Collins, Colorado 80523, USA*
[4] *Frontier Research System, The Institute of Physical and Chemical Research (RIKEN), Saitama 351-0198, Japan*

ABSTRACT

We monitor the stick-slip displacements of a very slowly driven moveable perforated top plate which interacts via shearing with a packing of identical glass beads confined in a tray. When driven at a constant stress rate, the distributions of large event displacements and energies triggered by the stick-slip instabilities exhibit power-law responses reminiscent of the Gutenberg-Richter law for earthquakes. Small events are quasi-size independent, signaling crossover from single bead transport to collective behavior.

*Introduction.—* There have been many theoretical studies (e.g., [1–5]) of the Gutenberg-Richter law [6–9] for the power-law rank-order distribution of earthquakes sizes $N(\geq m)$ which obeys $\text{Log}_{10}(N(\geq m)) \sim a - b\,m$. The seismic magnitude m is a logarithmic measure of event size with *b* in the range 0.80 to 1.06 for small events and 1.23 to 1.54 for very large ones [2]. In contrast to these many theoretical studies, only a few well-controlled laboratory experiments of sheared systems have generated event size and energy distributions [10, 11]. Thus, there is a clear need to do laboratory experiments to study these distributions in a variety of real sheared systems. Several spatial-temporal extended systems exhibit broad-tailed distributions of avalanches when very slowly driven to the threshold of instability. Examples include water droplet avalanches [12], vortices in superconductors [13], magnetic domain wall motion [14] and other systems [9,15]. In granular media undergoing prolonged grain bed ramping [16,17], gravity provided the shear force. Many careful experiments have explored the dynamic characteristics of slowly driven granular systems, such as the strong force fluctuation due to the formation of stress chains [18] and friction that results in stick-slip motion [19-21]. The statistics of events, however, was not the focus of these studies in sheared granular media. Only a single study, of sheared tapioca grains in an annular cell [11], observed a power-law distribution of responses.

We monitor the stick-slip displacements of a very-slowly-driven horizontal perforated plate which interacts via shearing with a packing of identical glass beads confined in a tray. When the top plate is pulled across the beads, we observe that the distributions of displacements triggered by the stick-slip instabilities exhibits a power-law response for large events, reminiscent of the Gutenberg-Richter law for earthquakes. Our experimental arrangement is distinctly different from [11]'s who used polydisperse tapioca grains (5 $\mu$m < $d$ < 2000 $\mu$m) with depths of 25 mm to 30 mm and a rotating annular plate to slowly shear the system. In our system, only three bead layers are used in order to generate the maximal shear. Glass beads of diameter d = 5 mm are slowly and linearly sheared in a quasi-two dimensional square cell, similar to that used in [19]. Our beads are confined and under compression in a cell shallower in terms of grain layers than [19]'s, which contained glass beads with 70 $\mu$m < $d$ < 110 $\mu$m and height ~ 2 mm, or ~ 20 layers in depth.

*Experiment.—* Our apparatus consists of a 23×25 cm$^2$ tray containing a plastic substrate and three separate layers of 0.167 gm, 5 mm diameter glass balls. The flat substrate (inset to Fig. 1c) presents a regular triangular array of about 2300 cup-shaped indentations, each ~ 2 mm deep and 5.5 mm in both diameter and lattice spacing. Because of good substrate commensurability the three layers of balls tend to be hexagonal close-packed. A top driving plate of identical corrugation to the substrate plate is placed above the ball layers (inset to Fig. 1c). It is secured to a 0.9 cm thick Plexiglas plate tied directly to an ultra-low speed motor via a length of high-strength braided nylon twine. A lid plate provides a compressive force on the top plate. Compression is applied with four small springs and adjusting screws inserted between the lid and frame. The typical force of about 20 N permits a few mm of top plate vertical motion. To ensure identical initial conditions for all runs the bed is prepared by crystallizing the tray beads. The nylon cord is then adjusted for fit, the upper plates reattached and each spring is compressed to the specified height before starting a run. Data are collected as the 46 cm drive plate travels through its entire 21 cm range in 2.5 hours. During runs the tractor motion of the drive plate rakes all three bead layers steadily forward against the front lip of the tray, and (to prevent jamming) frees the top bead layer over it (~30% of the beads are eventually freed per run).

Data collection uses a high-precision optical encoder to measure the top plate's displacement. Each digitized bit of the encoder corresponds to a displacement of about 10–15 $\mu$m. Whenever the plate moves by one bit, the time is recorded to an accuracy of about 5 µs. About 16,000 such times are recorded during a run, from which we can find the start times of displacement events; their sizes, energies, lifetimes, and waiting times between events.

*The Data.—* Six final runs were performed, each at low and high spring compressions. The resulting data were size-sorted into rank-ordered size distributions and histograms. The top driven plate exhibits a clear sequence of stick and slip events. In Fig. 1(a) a time sequence of plate velocities (~10% of one run, enlarged in (b), and (c)) and displacements (d) for the full run demonstrate a series of events ranging in size from 10 µm to 1 cm. The development of these motions is neither uniform in size nor periodic in time, but instead exhibits a broad distribution of different lengths and time intervals between events. The power versus frequency plot for two typical data sets, obtained by Fourier transforming the raw data, is presented in Fig. 2 (upper inset). The power-law of the data's power spectrum, $f^{-2.0}$, spans > 2.5 decades as determined by an eyeball fit. This is consistent with broad distributions found in other stick-slip [11] and avalanching experiments [10].

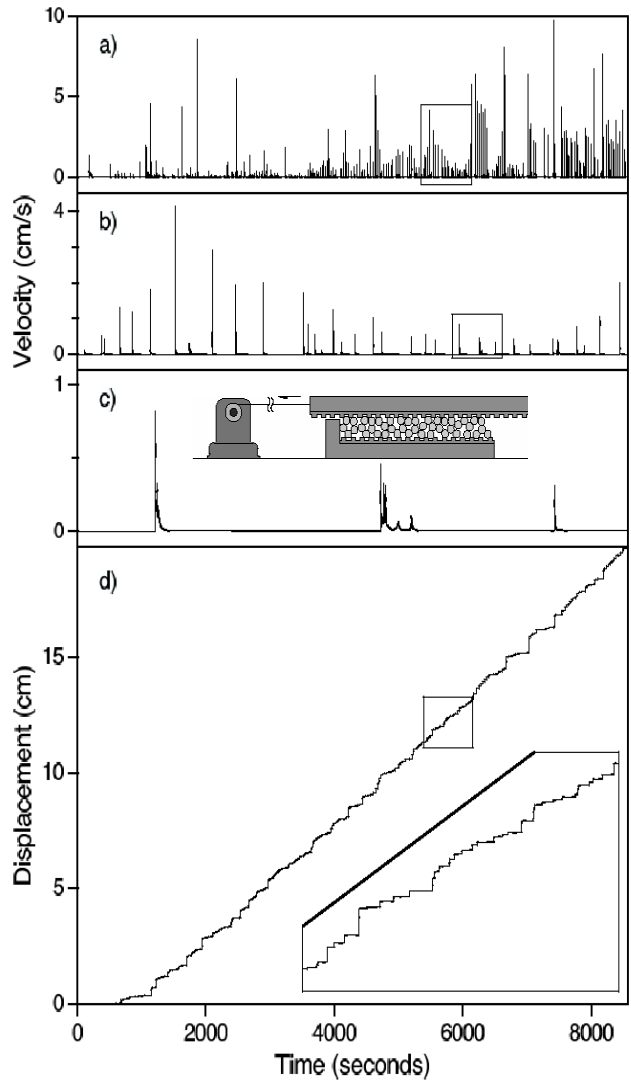

FIG. 1. Typical time series of the driving plate velocity (panels (a), (b), and (c)) and of the plate's full displacement (panel (d)). Small square in (a) is magnified in (b), similarly from (b) to (c), which contains schematic of setup. The inset at the bottom right shows the magnification of a small segment of the displacement versus time (inside the square in (d)). Setup shown in panel (c) inset.

*Event Size Analysis.–* The event size *s* is taken as the number of recorded data points that are bracketed by quiescent "threshold" time intervals, *T*. The threshold criterion corresponds to the maximum time allotted between each recorded data point assumed to belong to the same displacement event. Three different threshold times (0.05 s, 0.095 s, and 0.30 s) were chosen in constructing rank-ordered (RO) distributions of event size. These distributions, which show the number of events with size ≥ s, versus s, are compared in Fig. 2. Each curve follows a quasi-power-law behavior in the RO size distribution at large s, but have gentle roll-offs for small event sizes. The curves with T ≤ 0.300 s are nearly identical and have more extended power-law regions than do the higher T curves (not shown). This happens with increasing T because many smaller sequential events eventually become subsumed into a lesser number of larger events, reducing the power-law region and slope.

The data for T = 0.30 s are replotted in Fig. 2 (lower inset) as a histogram of the size distribution, $D(s)$. The power-law region is least squares fit by $D(s) = 3.77\, s^{-1.85}$ over 1.3 decades in s (dashed line). Since the histogram is roughly the derivative of the rank-ordering plot, the exponent in the fit of $D(s)$ can be represented as $-(b+1)$, or $b = 0.85 \pm 0.2$. Therefore, our *b* is consistent with the Gutenberg-Richter law for small events.

The roll-off region at small values of s in Fig. 2 proper corresponds to a deviation of the histogram from power-law behavior to a broad low-s plateau. The crossover from power-law to plateau indicates that an additional scale-setting parameter is present. We therefore least squares fit the histogram data with the modified power-law relation,

$$D(s) \sim (s + s_0)^{-(b+1)}. \tag{1}$$

This data fit extends over more than 1.5 decades in s (solid line of Fig. 2 inset), producing $b = 1.2 \pm 0.5$, and specifying a characteristic length scale, $s_0 \approx 18$, corresponding to a 180–270 μm displacement. This length scale reflects an underlying property of the glass bead bed. Since $s_0$ is 3 % – 6 % of a bead diameter, we suggest that this length represents either i), the effective relaxation distance of single-bead stick-slip events, in agreement with [19]'s localized rearrangements, or ii), the size beyond which events become fully 2D (constrained by the top and bottom plates of our three-bead-thick apparatus). Case ii) would be analogous to that of very large (subduction zone) earthquake distributions [2].

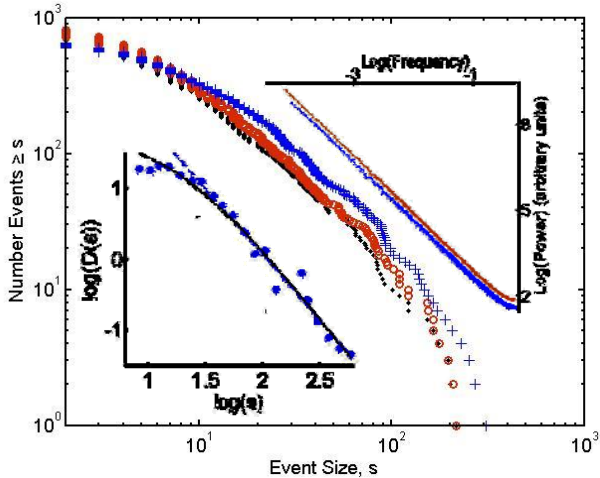

FIG. 2. (color online). Rank-ordered distributions of event size s for minimum event time separations T = 0.050(·), 0.095(o), 0.30(+) s). Upper inset: Power versus frequency plots for two data sets (f is relative to Nyquist $f_N = 0.5$). Lower inset: log-log plot of size distribution histogram, D(s) (in arbitrary units), using T = 0.30 s ( ● ). Dashed (solid) line is best fit of data to power-law (modified power-law) (see text).

*Event Energy Analysis*.– The velocity spikes of Fig. 1(a)-1(c), suggest that normalized kinetic energies, E, can be taken as the squares of *maximum* event velocities. All runs were combined into two large constant-compression data sets, and then reanalyzed using a threshold value of T = 0.095s. (The energy distributions are *insensitive* to T since only the maximum event velocity is kept, with smaller events simply subsumed.) Squaring of the velocity expands the dynamic range of measured events from three orders of magnitude in size to eight orders in energy. With this sensitivity enhancement, however, comes an amplification of systematics within each data run; in particular, event energies tend to grow in size as runs progress. Separate rank-ordered energy distributions were therefore constructed from the combined first halves and second halves of all data runs taken at low compaction and compared (not shown). These partial energy distributions themselves display identical power-law slopes, so we may proceed with the confidence that runs maintain adequate statistical stationarity.

The RO data curves (Fig. 3) for both low and high compression runs have plateaus at small E and power-law falloffs at higher energy, while the corresponding histograms possess maxima where the RO plateaus begin. Thus, a characteristic scale is again present in the distributions. Comparing the two RO curves reveals a crossover energy, $E_0$, below which low-compression events are more numerous, and above which high-compression events are more common. This trend is to be expected, as bead movement under light spring loading is relatively less constrained than at higher loading, but they also retain less pent-up energy for big events. A three-decade-long region of the low compression rank-order curve (○) is well represented by the power-law function (dot-dash line)

$$N(\geq E) = 3.5/(E + E_0) \approx E^{-1.0} \text{ for large E,} \qquad (2)$$

with $E_0 = 0.005$, while both low and high energy regions coincide with the high compaction rank-order distribution curve (● in Fig. 3). Our physical interpretation is that there exist two dynamical regimes which are best accessed by the higher compression runs. However, in the crossover region the high compression tends to spread the transition out and also to occasionally jam the system up (not shown). Conversely, a better formed RO knee is apparent at smaller compression, signaling a more distinct transition [22].

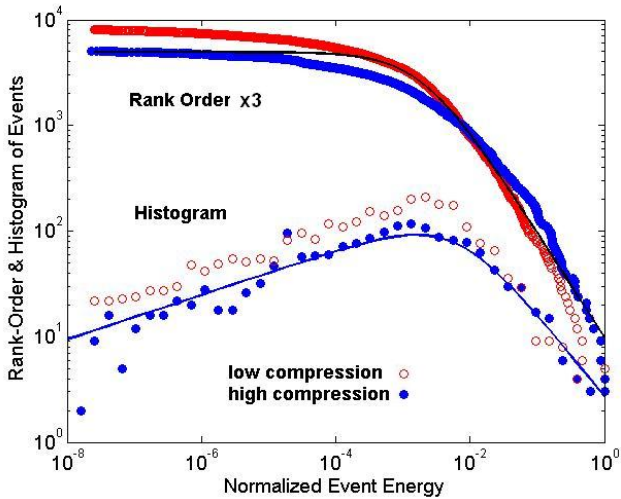

FIG. 3. (color online) Rank-order and Histogram distributions of normalized event energy, E, are shown for runs at low (○) and high (●) compression. Solid lines are modified power-law fits (see text). The RO curves have been offset for clarity.

The histogram for the high compression data set (●) follows $D(E) = 2E^{1/4}/(E + E_0)$ (dashed line). So, below $E_0 = .005$ the distribution is characterized by $D(E) \sim E^{1/4}$, an increasing function of E. Evidently, beads tend toward incommensuration with the indentation arrays on the lower and upper substrates as stresses build, resulting in a weakly E-dependent activity before crossover to collective behavior at $E_0$.

Finally, we compare our RO power-law exponent, 1.0 in Eq. (2), to the Gutenberg-Richter exponent recast in terms of total energy release[2], $N(\geq E) \sim E^{-b/d}$. For small earthquakes $d = 1$ so $b$ is 1, and for larger earthquakes $d = 3/2$ so $b$ would be 1.5. Thus, the energy dependent power-law falloff is consistent with both the small and large earthquake power-law distributions, $0.80 \leq b/d \leq 1.06$, and $1.23 \leq b/d \leq 1.54$, respectively [2].

Among the several open-container, gravity-driven granular media setups for exploring the dynamics of avalanching, small event deviations from power-law behavior are found for all of them. In addition, Rogers and Head [23] have suggested that a crossover can occur between distinct modes of transport within avalanche beds. In particular, they model avalanching rice pile dynamics as crossing over from rolling transport of individual grains to over-damped motion dominated by clustered grain motion associated with long-tailed distributions. In our closed, compressed system, the presence of scale parameters $s_0$ and $E_0$ suggest that the glass beads have two distinct modes of transport. They can either move singly, resulting in either D(s) plateaus or weakly energy-dependent behavior, or collectively, suffering stress network collapse that produces long-tailed distributions.

*Conclusions.—* We performed a table-top stick-slip experiment in which glass beads were sheared by a slowly-moving perforated plate. We analyzed both relative displacements and normalized peak energies of the instabilities and found the resulting power-law response to be comparable to that of stick-slip earthquakes. The analysis reveals a scale parameter marking the crossover from single grain movement to collective behavior.

*Acknowledgements.—* FN acknowledges partial support from the NSA and ARDA, under AFOSR contract F49620-02-1-0334 and also by the NSF grant No. EIA-0130383. Prof. Zaretzki acknowledges past NSF-REU support.

References

\*\* Present address: Dept. of Physics, Kyushu University.
[1] J.M. Carlson, J.S. Langer, and B.E. Shaw, Rev. Mod. Phys. **66**, 657 (1994), and references therein.
[2] Z. Olami, H.J.S. Feder, and K. Christensen, Phys. Rev. Lett. **68**, 1244 (1992); P. Bak, K. Christensen, L. Danon and T. Scanlon, ibid. **88**, 178501 (2002).
[3] P. Miltenberger, D. Sornette, and C. Vanneste, Phys. Rev. Lett. **71**, 3604 (1993); V. De Rubeis *et al*, *ibid.* **76**, 2599 (1996); J.B. Rundle, *et al.*, *ibid.* **76**, 4285 (1996); C.R. Myers, *et al.*, *ibid.* **77**, 972 (1996); D.S. Fisher, *et al.*, *ibid.* **78**, 4885 (1997); M.G. Shnirman and E. M. Blanter, *ibid.* **81**, 5445 (1998); D. Sornette and A. Helmstetter, *ibid.* **89**, 158501 (2002).
[4] B. Barriere and D.L. Turcotte, Phys. Rev. E **49**, 1151 (1994); H.J. Xu and L. Knopoff, *ibid.* **50**, 3577 (1994); H. Ceva, *ibid.* **52**, 154 (1995); M.E.J. Newman and K. Sneppen, *ibid.* **54**, 6226 (1996); J.B. Rundle, W. Klein, S. Gross, and C.D. Ferguson *ibid.* **56**, 293 (1996). R. Hallgass *et al*, *ibid.* **56**, 1346 (1997).

[5] H. Nakanishi, Phys. Rev. A **41**, 7086 (1990); H. Nakanishi, *ibid.* **46**, 4689 (1992). K. Chen, P. Bak, and S.P. Obukhov, *ibid.* **43**, 625 (1991); G.L. Vasconcelos, M. de Sousa Vieira, and S.R. Nagel, *ibid.* **44**, R7869 (1991); L. Knopoff, J.A. Landoni, and M.S. Abinante, *ibid.* **46**, 7445 (1992).

[6] B. Gutenberg and C.F. Richter, Ann. Geophys. **9**, 1 (1956).

[7] D.K. Turcotte, *Fractals and Chaos in Geology and Geophysics*, 2nd ed. (Cambridge, Cambridge, 1997).

[8] C.H. Scholz, *The Mechanics of Earthquakes and Faulting*, 2nd ed. (Cambridge, Cambridge, 2002).

[9] H.J. Jensen, *Self-Organized Criticality* (Cambridge University Press, Cambridge, 1988), especially Ch. 2.

[10] H.J.S. Feder and J. Feder, Phys. Rev. Lett. **66**, 2669 (1991); S. Ciliberto and C. Laroche, J. de Physique I **4**, 223 (1994).

[11] F. Dalton and D. Corcoran, Phys. Rev. E **63**, 061312 (2001); **65**, 031310 (2002).

[12] B. Plourde, F. Nori, and M. Bretz, Phys. Rev. Lett. **71**, 2749 (1993).

[13] S. Field, *et al.*, Phys. Rev. Lett. **74**, 1206 (1995); O. Pla and F. Nori, *ibid.* **67**, 919 (1991); R.A. Richardson, O. Pla and F. Nori, *ibid.* **72**, 1268 (1994); C.J. Olson, C. Reichhardt, and F. Nori, Phys. Rev. B **56**, 6175 (1997); C.J. Olson *et al.*, Physica C **290**, 89 (1997); E. Altshuler and T.H. Johansen, Rev. Mod. Phys. **76**, 471 (2004).

[14] G. Durin and S. Zapperi, Phys. Rev. Lett. **84**, 4705 (2000); S. Zapperi, *et al.*, Phys. Rev. B **58**, 6353 (1998).

[15] S. Field, N. Venturi, and F. Nori, Phys. Rev. Lett. **74**, 74 (1995).

[16] M. Bretz, J.B. Cunningham, P.L. Kurczynski, and F. Nori, Phys. Rev. Lett. **69**, 2431 (1992); G.A. Held *et al.*, *ibid.* **65**, 1120 (1990); J. Rosendahl, M. Vekic, and J. Kelley, Phys. Rev. E **47**, 1401 (1993).

[17] V. Frette *et al.*, Nature **379**, 49 (1996).

[18] R.R. Hartley and R.P. Behringer, Nature **421**, 928 (2003); D. Howell, R.P. Behringer, and C. Veje, Phys. Rev. Lett. **82**, 5241 (1999); B. Miller, C.O'Hern, and R.P. Behringer, Phys. Rev. Lett. **77**, 3110 (1996).

[19] S. Nasuno, A. Kudrolli, and J.P. Gollub, Phys. Rev. Lett. **79**, 949 (1997); S. Nasuno, A. Kudrolli, A. Bak, and J.P. Gollub, Phys. Rev. E **58**, 2161 (1998).

[20] J.-C. Geminard, W. Losert, and J.P. Gollub, Phys. Rev. E **59**, 5881 (1999); W. Losert, J.-C. Geminard, S. Nasuno, and J.P. Gollub, *ibid.* **61**, 4060 (2000).

[21] R. Albert, M.A. Pfeifer, A.-L. Barabasi, and P. Schiffer, Phys. Rev. Lett. **82**, 205 (1999); I. Albert *et al.*, *ibid.* **84**, 5122 (2000).

[22] R. Richard, *et al.*, Nature Materials **4**, 121 (2005).

[23] D.A. Head and G.J. Rodgers, J. Phys. A: Math. Gen. **32**, 1387 (1999).